\documentclass[9pt,twocolumn,twoside]{opticajnl}
\journal{opticajournal} 

\setboolean{shortarticle}{true}

\usepackage{lineno}

\title{Replica-assisted super-resolution fluorescence imaging in scattering media}

\author[1,2]{Tengfei Wu}
\author[1]{YoonSeok Baek}
\author[1]{Fei Xia}
\author[1]{Sylvain Gigan}
\author[1,*]{Hilton B de Aguiar}
\affil[1]{Laboratoire  Kastler  Brossel,  ENS-  Université  PSL,  CNRS,  Sorbonne  Université,  Collège de France. 24 rue Lhomond, 75005 Paris, France}
\affil[2]{State Key Laboratory of Transient Optics and Photonics, Xi’an Institute of Optics and Precision Mechanics, Chinese Academy of Sciences, 710119 Xi’an, China}
\affil[*]{Corresponding author: h.aguiar@lkb.ens.fr}

\begin{abstract}
Far-field super-resolution fluorescence microscopy has been rapidly developed for applications ranging from cell biology to nanomaterials. However, it remains a significant challenge to achieve super-resolution imaging at  depth in opaque materials. In this study, we present a super-resolution microscopy technique for imaging hidden fluorescent objects through scattering media, started by exploiting the inherent object replica generation arising from the memory effect, i.e. the seemingly informationless emission speckle can be regarded as a random superposition of multiple object copies. Inspired by the concept of super-resolution optical fluctuation imaging, we use  temporally-fluctuating speckles to excite fluorescent signals and perform high-order cumulant analysis on the fluctuation, which can not only improve the image resolution, but also increase the speckle contrast to isolate only the bright object replicas. A super-resolved image can be finally retrieved by simply unmixing the sparsely distributed replicas with their location map. This methodology allows to overcome scattering and achieve robust super-resolution fluorescence imaging, circumventing the need of heavy computational steps.
\end{abstract}

\setboolean{displaycopyright}{false} 

\begin{document}

\maketitle

\section{Introduction}
Fluorescence microscopy is the workhorse in bio-imaging, with its resolution limited by diffraction. Many approaches have been proposed and successfully achieved super-resolution far-field fluorescence imaging\cite{Huang2010,Schermelleh2019}, using optical non-linearity to shrink the probe size\cite{Hell1994}, computational methods to retrieve high-spatial-frequency information outside the cutoff frequency of the optical system using structured illumination\cite{Gustafsson2000}, source localization schemes\cite{Betzig2006} or stochastic reconstruction\cite{Rust2006,Dertinger2009}. Although current methods can properly address the challenge of surpassing classical imaging resolution, they are mostly limited to imaging transparent samples or only the superficial region of the optically thick media: they are limited by optical scattering.

To image deeper in super-resolution microscopy, adaptive optics has been exploited to mitigate the issue of image degradation at depth\cite{Burke2015,Velasco2021,Lin2021}. However, the adaptive optics method is designed for aberration correction, where aberrations come mainly from ballistic photons and are dominant by limited low-order Zernike modes\cite{Booth2015}. Therefore, it is insufficient for imaging at greater depth where multiple scattering occurs. In the past decade, wavefront shaping\cite{Vellekoop2007,Popoff2010} has emerged to overcome multiple scattering\cite{Rotter2017,Yoon2020,Horstmeyer2015}. Compared to adaptive optics, wavefront shaping methods can handle much higher complexity, enabling focusing or imaging at depth. However, the application of super-resolution fluorescence imaging in the multiple-scattering regime using wavefront shaping may require special guide-stars with subwavelength sizes\cite{Kim2019}, to provide the interior feedback to push the light beam to focus below the diffraction limit, which remains difficult. Alternatively, computational imaging concepts have recently been exploited to see hidden targets through scattering media\cite{Bertolotti2022}. The great majority of these computational methods exploit speckle correlations\cite{Bertolotti2012,Katz2014}, the so-called optical memory effect\cite{Feng1988,Freund1988}: shifting/tilting the incoming wavefront would induce only a global and lateral speckle shift, instead of changing the speckle structure (up to a certain limit). Thanks to the memory effect, the captured image can be modeled as an intensity convolution of the object image and the shift-invariant speckle. Using this property, several algorithms, such as phase-retrieval\cite{Fienup1982}, bispectrum analysis\cite{Wu2016}, or deconvolution\cite{Edrei2016,Wu2020}, can reconstruct a hidden object.

\begin{figure*}[ht]
\centering
\fbox{\includegraphics[width=0.9\linewidth]{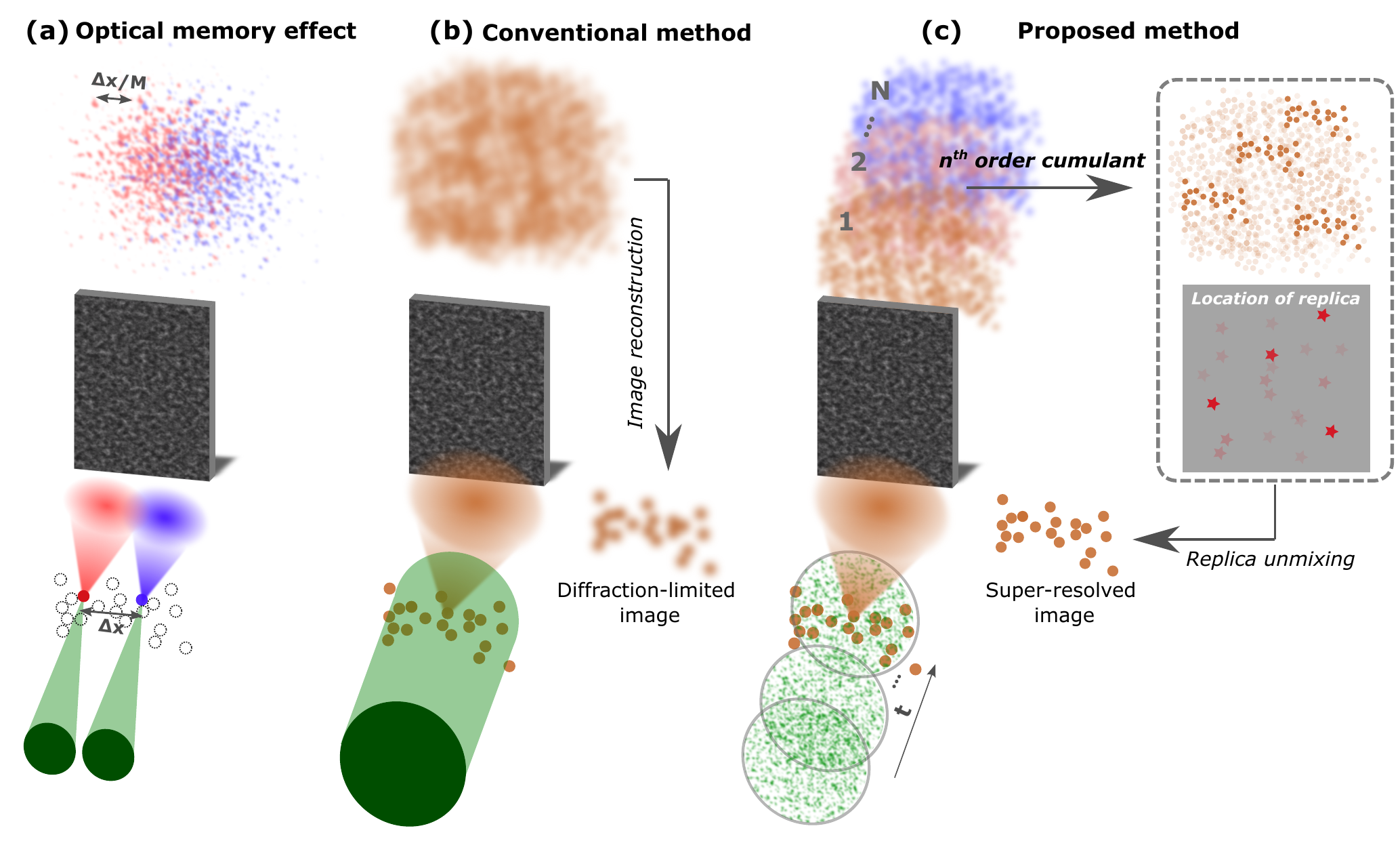}}
\caption{Principle of super-resolution imaging through scattering media exploiting object replicas and cumulant analysis. (a) Illustration of optical memory effect. Two point sources of the hidden object with a distance ${\Delta x}$ are excited. When the two points locate within the memory effect, two identical speckle patterns with a lateral translation $\frac{\Delta x}{M}$ are generated, overlapping incoherently. (b) Upon uniform excitation of the whole object, many object replicas, locating at the position of speckle grains, are randomly superimposed, generating a low-contrast speckle pattern. In this conventional method, only a diffraction-limited image could be reconstructed from the recorded camera image. (c) In the proposed method, uncorrelated speckles excite the object, and a stack of temporally fluctuating speckle patterns are recorded, which are used to calculate the $n^{th}$ order cumulant speckle image. The replicas in the cumulant speckle image are super-resolved and the remaining bright ones are isolated. A location map of the replicas is created and used to unmix the replicas to obtain the final super-resolved object image.}
\label{fig:principle}
\end{figure*}

Recent efforts for deep super-resolution imaging have utilized several of these concepts. The exploitation of speckle correlation through the memory effect has enabled the reconstruction of super-resolved images. This is achieved by harnessing the blinking of fluorophores excited by speckles in conjunction with the principles of single-molecule localization\cite{Wang2021}. However, this method requires the object excited in each frame to be sparsely distributed, limiting its wider applications. An improved version of this method has been proposed\cite{Zhu2024}, in which a stack of fluctuating speckle realizations is collected. A methodology based on phase-retrieval algorithm and high-order cumulant super-resolution method \cite{Dertinger2009} is used to retrieve a super-resolved image. However, this method is not ideal for low signal-to-noise (SNR) conditions, which is often encountered in fluorescence imaging through scattering media. In addition, phase-retrieval is an ill-posed problem and highly sensitive to statistical noise of speckle autocorrelation, which determines that the iterative process may converge to the local minima and suffer from the ambiguity problem.

To circumvent these issues, we propose a robust super-resolution fluorescence microscopy that achieves super-resolution imaging in scattering media using a simple computational framework. Our method starts with the memory effect of scattering, wherein the intensity speckles generated by the point sources are highly correlated, and the full speckle pattern can be regarded as a random collection of numerous object replicas. Inspired by the classic technique of super-resolution optical fluctuation imaging (SOFI)\cite{Dertinger2009}, we induce temporally fluctuating signal by exciting the fluorescence with random speckle illuminations and record a stack of fluorescent speckles. A high-order cumulant analysis is then performed on the recorded speckle realizations, offering two significant benefits thanks to the nonlinearity of cumulant calculation. On the one hand, it can improve the image resolution, and on the other hand, it can greatly increase the speckle contrast: increasing the speckle contrast can effectively suppress dim replicas, making the speckle image much less complex. A location map and the weight of the remaining replicas in the cumulant speckle image can be identified by using a cross-correlation calculation. The final super-resolved image can be simply retrieved through an unmixing procedure, i.e. deconvolving the high-order cumulant speckle image with the location map. Compared to speckle-correlation based super-resolution imaging methods, our method is more computationally straightforward, and more robust for fluorescence microscopy since we avoid using the iterative phase-retrieval process and do not need to estimate the full complex speckle. Instead, we work only with the bright replicas in the cumulant speckle image.

\section{Principle}
\subsection{Object replicas in the camera image}

Fig. \ref{fig:principle} illustrates the basic principle of our method. We start describing the so-called optical memory effect\cite{Feng1988,Freund1988}, where a wavefront tilt of the beam would only induce a lateral global shift in the intensity of scattered light. We consider a hidden object located within the region determined by the memory effect, and two point sources on the object, separated by $\Delta$x, are excited and emit fluorescence. Since the sources are within the memory effect range, the generated two fluorescent speckles are nearly identical, and incoherently superimposed with a relative translation of $\frac{\Delta x}{M}$, where \emph{M} is the magnification of the optical system (Fig. \ref{fig:principle}(a)). Similarly, excitation of the full hidden object gives a low-contrast speckle pattern (Fig. \ref{fig:principle}(b)), resulting from the superposition of many identical fluorescent speckles with certain translations. Therefore, the resulted low-contrast speckle pattern can be considered as a collection of multiple superimposed replicas, locating at the position of the speckle grain and with various brightness. Mathematically, the low-contrast speckle pattern can be described as a convolution of the hidden object and a shift-invariant point-spread function (PSF), which is the fluorescent speckle from a single point source\cite{Katz2014}:
\begin{equation} \label{eq:1}
I(r) = O(r) \ast S(r)
\end{equation}
where ``$\ast$" is the convolution operator. \emph{I}(r) denotes the camera image, \emph{O}(r) and \emph{S}(r) are the hidden object and the speckled PSF. Due to the complex superposition of object replicas, it is always challenging to identify the object image from the seemingly information-less speckle, as shown in Fig. \ref{fig:principle}(b). Additional computational steps, such as phase-retrieval, are generally required to reconstruct the final image, which however is diffraction-limited.

\subsection{Replica suppression and resolution improvement by optical fluctuation}
With the understanding that the complex speckle pattern on the camera is the collection of randomly superimposed object replicas, an object image can theoretically be retrieved by simply shifting these replicas and aligning their centers\cite{Hwang2023}. However, the object replicas can not be easily identified due to the low contrast of the speckle pattern, especially for the object with complex structure. Due to the random nature of speckle, the brightness of speckle grains in $S$ of Eq. \ref{eq:1} has significant variation, as regards the object replicas because of the linearity of convolution. Therefore, increasing the speckle contrast can properly suppress the dim replicas and make the camera image much less complex for computation.

\begin{figure*}[ht]
\centering
\fbox{\includegraphics[width=0.9\linewidth]{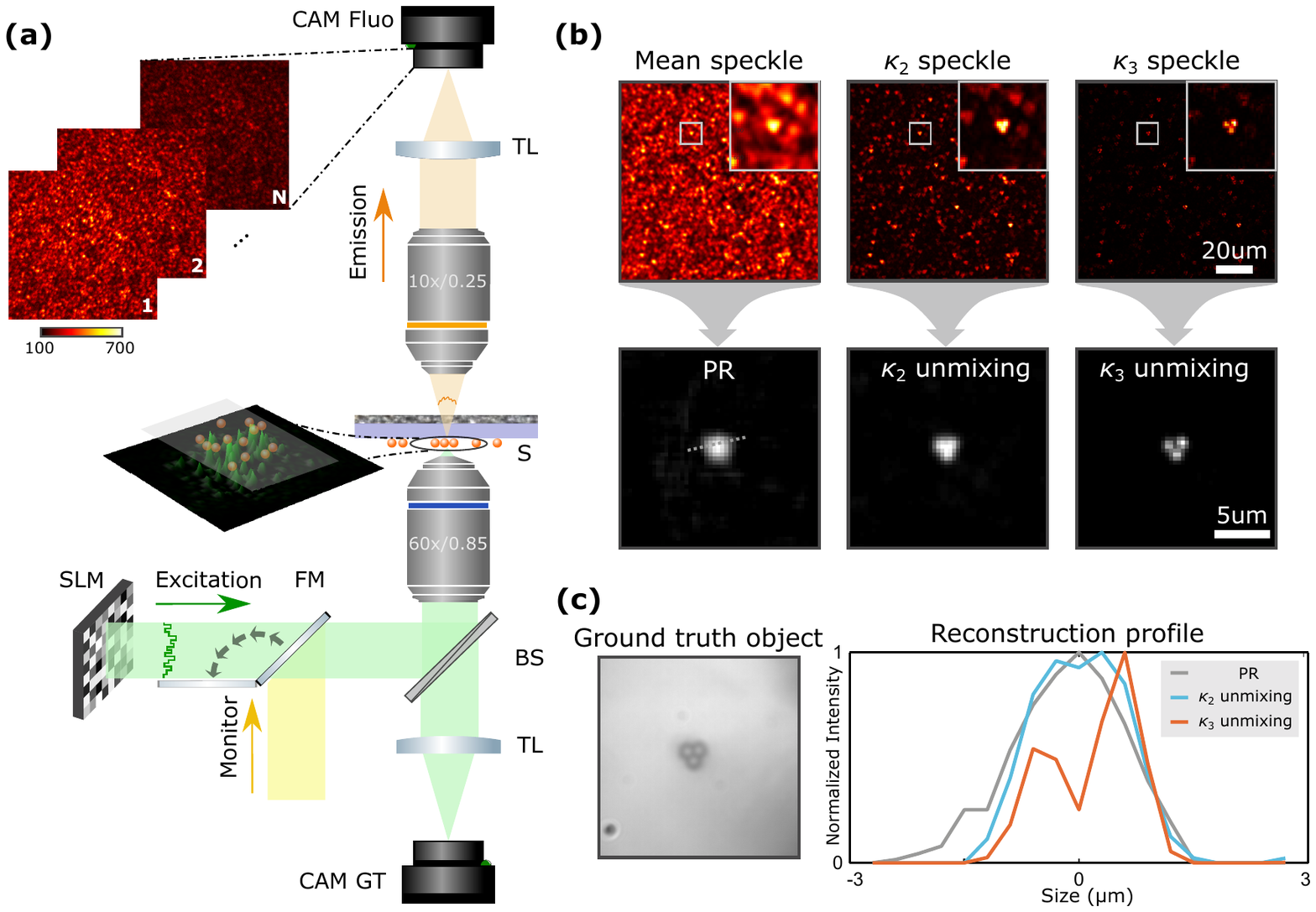}}
\caption{Experimental setup and proof-of-concept results. (a) Experimental setup. A series of random phase patterns are shown on the SLM and the corresponding uncorrelated speckle illuminations are generated on the sample plane (S) through a beamsplitter (BS) and a high-NA (0.85) objective to excite the fluorescent beads. The emitted fluorescent signal are transmitted through a scattering medium and collected with an objective (NA=0.25), and then imaged on CAM fluo with a tube lens (TL). The top inset shows representative recorded fluorescent speckle patterns. A flip mirror (FM) is used to switch the illumination between fluorescence excitation laser and the white-light, using for recording the ground truth object with CAM GT. (b) Top row: Mean speckle, $\kappa_{2}$ and $\kappa_{3}$ cumulant speckle images. The insets compare the cropped object replica in respective speckle images. Bottom row: diffraction-limited image from the mean speckle, processed by phase-retrieval algorithm, and the super-resolved results by, respectively, unmixing the replicas in $\kappa_{2}$ and $\kappa_{3}$ cumulant speckle images. (c) Ground truth object, and a comparison of the profiles in panel (b).}
\label{fig:EXP_Proof_of_concept}
\end{figure*}

To decrease the density of the object replicas and robustly retrieve the object image, we use the high-order cumulant calculation of a stack of temporally fluctuating speckle patterns. Instead of using uniform excitation in the conventional method, as shown in Fig. \ref{fig:principle}(b), a series of random speckles continuously excite the fluorescent object, inducing a temporally optical fluctuation (Fig. \ref{fig:principle}(c)), and the corresponding measurement for each realization can be expressed as follows:
\begin{equation} \label{eq:2}
I(r,t) = [O(r) \times P(r,t)] \ast S(r)
\end{equation}
where $P(r,t)$ denotes the excitation speckle at time $t$. Assuming that the grain size of the excitation speckle is sufficiently small, the fluctuation of any position within an object can be regarded as being independent from the others. Therefore, the final $n^{th}$ order cumulant speckle image is written as (see Supplement 1, S1):
\begin{equation} \label{eq:3}
I_{{\kappa}_{n}} \propto O^{n}  \ast S^{n}.
\end{equation}

Eq. \ref{eq:3} shows how the high-order cumulant speckle image $I_{{\kappa}_{n}}$ preserves information about the object. Instead of convolving with $S$ (Eq. \ref{eq:1}), the object in $I_{{\kappa}_{n}}$ convolves with $S^{n}$, a non-linear speckled PSF. There are two important benefits from the cumulant images: i) it statistically improves the image resolution of the replicas by a factor of $\sqrt{n}$ \cite{Dertinger2009}; ii) it significantly increases the speckle contrast, concomitantly isolating the bright replicas in the camera image, making the speckle image sparse (Fig. \ref{fig:principle}(c)). 

\subsection{Reconstruction from high-order cumulant speckle image}
Because of the remaining bright replicas, the resulting cumulant speckle image is still ambiguous to correctly identify the hidden object. Therefore, we propose an unmixing procedure to deterministically extract the object from the cumulant speckle image without additional computational steps. Due to the fact that the remaining bright replicas are sparsely distributed in the high-order speckle image, we may rewrite the cumulant image as the following form (see Supplement 1, S1 for details):
\begin{equation} \label{eq:4}
I_{{\kappa}_{n}} \propto O^{n}  \ast S^{n} = \left ( O^{n} \ast h^{n} \right ) \ast \left ( \sum_{i}^{}\delta \left ( r-r_{i} \right )\times w_{i} \right )
\end{equation}
where $h^{n}$ denotes the $n^{th}$ order speckle grain. $\delta(r-r_{i})$ indicates the location of each replica and $w_{i}$ is the corresponding weights. According to Eq. \ref{eq:4}, the super-resolved object image $O^{n}$$\ast$$h^{n}$ can be retrieved simply by deconvolving the high-order cumulant speckle image $I_{{\kappa}_{n}}$ with the location map of the replicas (containing both the weights $w_{i}$ and the locations, see Fig. \ref{fig:principle}(c)), without any heavy computational steps. Thanks to the isolation of bright replicas in the high-order cumlant speckle, the location map can be identified with a well-established image processing method, such as cross-correlation (See more information below and in Supplement, S2). It is worth noting that the random local interference of scattering light makes each speckle grain vary in shape, which would imply a spatially-fluctuating convolution kernel. However, we assume in Eq. \ref{eq:4} that the high-order speckle image leads to identical kernels because of the nonlinearity in $h^{n}$ that effectively suppresses sidelobes.

\begin{figure*}[ht]
\centering
\fbox{\includegraphics[width=0.9\linewidth]{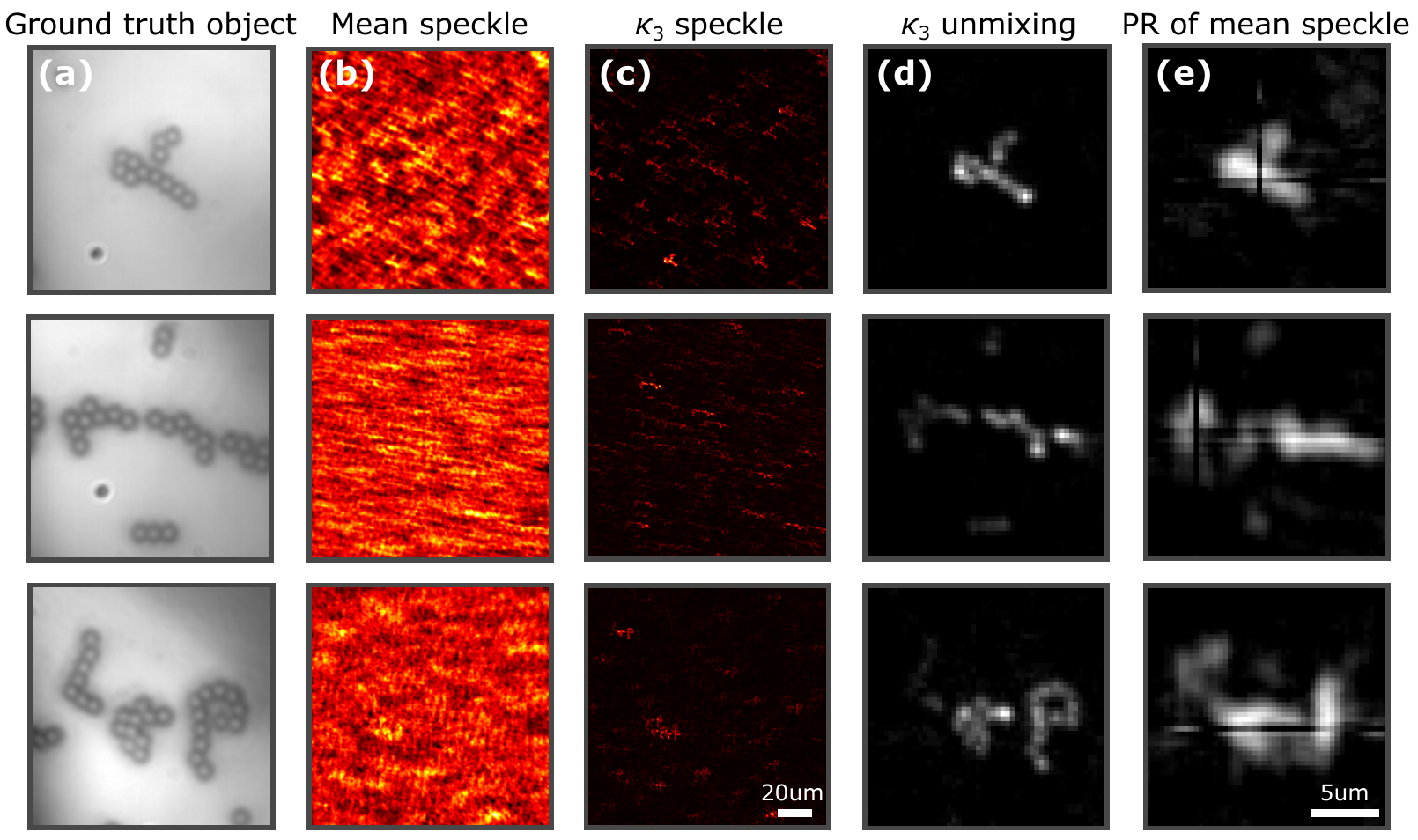}}
\caption{Replica-assisted super-resolution imaging of complex objects. (a) Ground truth objects. (b) Mean value of the temporally fluctuating fluorescent speckle patterns. (c) $\kappa_{3}$ cumulant speckle image. (d) Super-resolved image by unmixing the replicas in $\kappa_{3}$ speckle image. (e) Diffraction-limited image from the mean speckle with the phase-retrieval algorithm.}
\label{fig:EXP_complex_beads}
\end{figure*}

\section{Results}
\subsection{Experimental setup}
Fig. \ref{fig:EXP_Proof_of_concept}(a) shows the experimental setup of a wide-field fluorescence microscope for a proof-of-concept. A 532nm CW laser is used to excite the fluorescent signal from the sample consisting of 1$\mu$m orange fluorescent beads (540/560, FluoSpheres™), dropcasted on one side of a 1-mm-thick microscope slide. We sandblast the other side to induce a single scattering layer with optical memory effect (the spatial-correlation range is measured in Supplement 1, S3). A spatial light modulator (SLM) (HSP1920-532-HSP8, Meadowlark Optics) is used and various random phase patterns are sequentially sent to the SLM to generate multiple speckle realizations at the object plane. The modulated excitation is focused by an objective (CFI Plan Fluor 60XC, Nikon) with a numerical aperture (NA) of 0.85. A beamsplitter with a ratio 2/98 (transmission/reflection) is used to monitor the excitation speckle illuminations and to inspect the ground-truth with free-space brightfield imaging with a CMOS camera (acA4024-29um, Basler) and a tube lens (TL, f=150mm). The emitted fluorescent speckle patterns are collected by an objective (Plan N, Olympus) with NA=0.25 (nominal magnification 10X), and recorded on a sCMOS camera (Flash v4.0, Hamamatsu) with a TL (f=300mm). A set of bandpass filter (FBH560-10, Thorlabs), longpass filter (FELH0550, Thorlabs) and notch filter (NF533-17, Thorlabs) are used to sufficiently isolate the fluorescence signal.

\subsection{Proof-of-concept result}
We start by using a simple three-beads sample to demonstrate the speckle contrast enhancement and the resolution gain with cumulant analysis. The distance between adjacent beads is 1$\mu$m, which cannot be resolved (see wide-field imaging result in Supplement 1, S4) according to the Rayleigh criterion, that is, the diffraction-limit of the detection path is $\frac{0.61\lambda}{NA}\approx$1.37$\mu$m. By loading the random phase patterns on the SLM, the corresponding speckle patterns are generated on the sample plane to excite the fluorescence (inset of Fig. \ref{fig:EXP_Proof_of_concept}(a)) with adjustable field-of-view (FOV), and the speckle grain size is determined by the excitation NA. After 600 speckle realizations taken at 10fps, we calculate the mean value of the speckle realizations and the $\kappa_2$ and $\kappa_3$ cumulant speckle images (see more information in Supplement 1, S2), and the results are respectively shown in the first row of Fig. \ref{fig:EXP_Proof_of_concept}(b). We can clearly observe that the higher the order of the cumulant speckle images, the higher the contrast, thanks to the cumulant process. The dim replicas in the mean speckle are suppressed and only the remaining bright ones are sparsely distributed in the cumulant images. Furthermore, we compare one replica located at the same position of the three speckle images (insets of Fig. \ref{fig:EXP_Proof_of_concept}(b)), and it shows that the replica resolution is also increased with the cumulant process.

To extract the final object from the cumulant speckle image, we now describe a robust replica-unmixing process. Thanks to the fact that only the bright object replicas are sparsely distributed in the cumulant speckle image, according to Eq. \ref{eq:4}, we can unmix the object replicas using their "location map", $\sum_{i}^{}\delta \left ( r-r_{i} \right )\times w_{i}$, without additional heavy computational steps. Specifically, from both $\kappa_2$ and $\kappa_3$ cumulant speckle images, we select a reference sub-image containing the brightest object replica (See more information in Supplement 1, S2), with its size larger than twice the object image (determined by the speckle autocorrelation\cite{Wu2016}). To build the replica location map, first, the full high-order cumulant speckle image is converted into a stack of sub-images with the same size as the reference. Second, the cross-correlation between the reference and sub-images is calculated, which can quickly identify both the localization and the amplitude of the corresponding replicas. The final reconstruction, as shown in the second row of Fig. \ref{fig:EXP_Proof_of_concept}(b), is obtained with an unmixing step by deconvolving the high-order cumulant speckle image with the resulted location map (See Eq. \ref{eq:4} and more detailed procedures in Supplement 1, S2).

We compare the performance of high-order cumulant analysis with conventional method for imaging through scattering media. To achieve that, we use a phase-retrieval algorithm\cite{Bertolotti2012,Katz2014,Fienup1982} to process the mean speckle (equivalent to the speckle image in Fig. \ref{fig:principle}(b)) to retrieve the object image. Due to the inherent ambiguity of the phase-retrieval algorithm, the reconstruction process is repeated 200 times with different initial guesses, and the optimal one is used as the final result. Furthermore, we plot the profile of two adjacent beads of the reconstructions from different cases to compare the resolution (Fig. \ref{fig:EXP_Proof_of_concept}(c)). Compared to the diffraction-limited phase-retrieval image, both $\kappa_2$ and $\kappa_3$ reconstructions can improve the resolution, and $\kappa_3$ clearly resolves two adjacent beads, overcoming the diffraction-limit.

Finally, we demonstrate the robustness of the method for complex objects. Fig. \ref{fig:EXP_complex_beads} shows results of hidden objects with more complex structure composed by several connected fluorescent beads, together with their ground truth images shown in Fig. \ref{fig:EXP_complex_beads}(a). The largest sample is around 15$\mu$m, well locating within the memory effect range. 1000-2500 frames of fluorescent speckle patterns are recorded for the reconstructions, depending on the object complexity. Fig. \ref{fig:EXP_complex_beads}(b) shows the mean value of the measurements for each object. The low contrast of the mean speckle pattern challenges recognition of any detailed information of the hidden object. Conversely, the $\kappa_3$ cumulant speckle image in Fig. \ref{fig:EXP_complex_beads}(c) significantly suppresses the dim replicas, making the remaining ones in the full speckle image relatively sparse, with each replica better resolved than the corresponding one in the mean speckle (Fig. \ref{fig:EXP_complex_beads}(b)). Using the unmxing procedure, we retrieve the super-resolved image of each object in Fig. \ref{fig:EXP_complex_beads}(d), in which the adjacent beads are clearly resolved, overcoming the diffraction-limit of the imaging system. As comparison, Fig. \ref{fig:EXP_complex_beads}(e) shows the corresponding reconstruction from the phase-retrieval process from the mean speckle in Fig. \ref{fig:EXP_complex_beads}(b). We note strong artefact appearing in each reconstruction from phase-retrieval process, especially for the last two more complex cases. This observation is coherent with the fact that a faithful reconstruction from phase-retrieval process is related to a high-quality Fourier amplitude estimation from the speckle autocorrelation, requiring sufficient speckle grains (or object replicas) to suppress the statistical noise.

\section{Conclusion and discussion}
To conclude, we demonstrate a novel robust method for super-resolution imaging of fluorescent hidden objects through highly scattering media. Thanks to the optical memory effect and the linearity of intensity convolution operation, the recorded speckle pattern can be viewed as a random collection of many superimposed replicas of the object image. By exciting the temporally fluctuating fluorescent signal with random and uncorrelated speckle illuminations, the high-order cumulant analysis can effectively increase the replica resolution and speckle contrast, isolating many bright object replicas, sparsely distributed in the cumulant speckle image. The super-resolved image can finally be extracted by unmixing the remaining replicas in the high-order cumulant speckle image with the corresponding location map.

Our method is straightforward in terms of both implementation and computation. Regarding implementation, the linear optical excitation generates the fluorescent signal, allowing the use of wide-field imaging configuration, reducing the complexity of the microscope. It is worth noting that since the grain size of the excitation speckle is diffraction-limited, the fluctuation of adjacent emitters may still be correlated, making the resolution gain lower than the theoretical value of $\sqrt{n}$\cite{Kim2015,Choi2022}, found in classic SOFI, which assumes that the fluctuations of emitters are fully independent\cite{Dertinger2009}. In Supplement 1, S1, we mathematically analyze how the resolution gain is affected by the speckle grain size. We find numerically that the resolution can still be improved, even when the speckle grain size is comparable to the distance of adjacent emitters. To achieve optimal resolution gain, near-field speckle excitation\cite{Choi2022} with wide-field illumination, or side/tilted illumination excitation\cite{Laforest2020} can be a solution. In terms of computation, our method does not need prior knowledge of the low-resolution version of the hidden object\cite{Yilmaz2015,Zhu2024} or the PSF of the scattering imaging system\cite{Zhu2024}, thus avoiding ill-posed deconvolution and the risk of local minima in phase-retrieval process. Besides being straightforward, our data processing can be robust in noisy conditions (for instance the fluorescent signal through strong scattering), thanks to two facts: first, there is no complicated iterative process, and second, the nonlinear response of cumulant analysis helps to isolate only the bright (high SNR) replicas, contributing for the final reconstruction. Moreover, as the reconstruction does not need to suppress statistical noise, as in phase-retrieval-based approaches, our method has relatively relaxed constrain on the amount of speckle grains.

Besides the proof-of-concept demonstration in this work, there are several exciting extensions for our method, worth further being explored. First, our method is modeled based on the assumption of optical memory effect which limits the imaging FOV. A possible extension regarding this issue is to exploit the local memory effect\cite{Chen2022}, which allows to reconstruct hidden object within certain region, and a proper image stitching gives a large FOV. Second, an accurate location map is essential to properly unmix the object replicas in the cumulant speckle image for faithful reconstruction, especially in the highly noisy condition. Instead of using cross-correlation and maximum-value localization, a better strategy may be found to further improve the accuracy of the location map.

\begin{backmatter}
\bmsection{Funding} Agence Nationale de la Recherche (ANR-21-CE42-0013); H2020 Future and Emerging Technologies (863203); National Research Foundation of Korea (2022R1A6A3A03072108); HORIZON EUROPE Marie Sklodowska-Curie Actions (101105899).

\bmsection{Acknowledgments} The authors thank Thomas Chaigne for stimulating
discussions, and Lei Zhu for his valuable help in providing the codes of hardware control for data acquisition. Fei Xia acknowledges the funding from the Optica Foundation Challenge Award 2023. Sylvain Gigan acknowledges support from Institut Universitaire de France.

\bmsection{Disclosures} The authors declare no conflicts of interest.

\bmsection{Data availability} Data underlying the results presented in this paper are not publicly available at this time but may be obtained from the authors upon reasonable request.

\bmsection{Supplemental document}
See Supplement 1 for supporting content. 

\end{backmatter}

\bibliography{ref_main}



\end{document}